\documentclass[twocolumn,prl,showpacs,preprintnumbers,amsmath,amssymb,superscriptaddress]{revtex4-2}
\usepackage{graphicx}
\usepackage{color}
\usepackage{float}

\newcommand{\beq}{\begin{equation}}
\newcommand{\eeq}{\end{equation}}
\newcommand{\beqn}{\begin{eqnarray}}
\newcommand{\eeqn}{\end{eqnarray}}

\newcommand{\beginsupplement}{%
        \setcounter{table}{0}
        \renewcommand{\thetable}{S\arabic{table}}%
        \setcounter{figure}{0}
        \renewcommand{\thefigure}{S\arabic{figure}}%
     }

\begin{document}
\begin{titlepage}
This manuscript has been authored by UT-Battelle, LLC under Contract No. DE-AC05-00OR22725 with the U.S. Department of Energy. The United States Government retains and the publisher, by accepting the article for publication, acknowledges that the United States Government retains a non-exclusive, paid-up, irrevocable, world-wide license to publish or reproduce the published form of this manuscript, or allow others to do so, for United States Government purposes. The Department of Energy will provide public access to these results of federally sponsored research in accordance with the DOE Public Access Plan(http://energy.gov/downloads/doe-public-access-plan).
\newpage
\end{titlepage}

\title{Coexistence of symmetry-protected topological order and N$\rm\acute{e}$el order in the spin-1/2 ladder antiferromagnet C$_9$H$_{18}$N$_2$CuBr$_4$}
\author{Tao Hong}
\email[]{hongt@ornl.gov}
\affiliation{Neutron Scattering Division, Oak Ridge National Laboratory, Oak Ridge, Tennessee 37831, USA}
\author{Imam Makhfudz}
\affiliation{IM2NP, UMR CNRS 7334, Aix-Marseille Universit$\acute{e}$, Marseille 13013, France}
\author{Xianglin Ke}
\affiliation{Department of Physics and Astronomy, Michigan State University, East Lansing, Michigan 48824-2320, USA}
\author{Andrew F. May}
\affiliation{Materials Science and Technology Division, Oak Ridge National Laboratory, Oak Ridge, TN 37831, USA}
\author{Andrey A. Podlesnyak}
\affiliation{Neutron Scattering Division, Oak Ridge National Laboratory, Oak Ridge, Tennessee 37831, USA}
\author{Daniel Pajerowski}
\author{Barry Winn}
\affiliation{Neutron Scattering Division, Oak Ridge National Laboratory, Oak Ridge, Tennessee 37831, USA}
\author{Merc$\grave{\rm e}$ Deumal}
\affiliation{Departament de Ci$\grave{e}$ncia de Materials i Qu\'{\i}mica F\'{\i}sica $\&$ IQTCUB, Facultat de Qu\'{\i}mica, Universitat de Barcelona, Mart\'{\i} i Franqu$\grave{e}$s 1, Barcelona, Spain E-08028}
\author{Yasumasa Takano}
\affiliation{Department of Physics, University of Florida, Gainesville, Florida 32611-8440, USA}
\author{Mark M. Turnbull}
\affiliation{Carlson School of Chemistry and Biochemistry, Clark University, Worcester, Massachusetts 01610, USA}

\date{\today}

\begin{abstract}
Topological phases of matter are beyond the paradigm of Landau’s symmetry breaking and have challenged our understanding of condensed matter systems. Here we report a new type of symmetry-protected topological phase of matter in the spin-1/2 coupled two-leg ladder antiferromagnet C$_9$H$_{18}$N$_2$CuBr$_4$, DLCB for short. In this two-sublattice antiferromagnet with a weak easy-axis anisotropy, we find no evidence of a conventional spin-flop transition in the magnetization with the magnetic field applied parallel to the easy axis at \emph{T}=0.4 K, well below $T_{\rm N}$=2.0 K. Moreover, the temperature dependence of the gapped transverse excitations across $T_{\rm N}$ indicates that they are not the conventional \emph{S}=1 magnons associated with explicit symmetry breaking. Instead, the thermal renormalization of the gap energy shows a remarkable agreement with a calculation for the three-dimensional \emph{O}(3) nonlinear $\sigma$ model. Accordingly, the spin gap in DLCB is not due to the spin anisotropy but to the separation between a spin singlet state and a triplet excited state. Since an antiferromagnetic spin-1/2 ladder systems can be mapped onto the spin-1 chain, the notion of the Haldane gap is proposed to explain the opening of the spin gap in DLCB. Therefore, the ground state of DLCB is best described as a quantum superposition of a Haldane phase and a N$\rm\acute{e}$el-ordered phase, which resembles the quantum state of a qubit in quantum computing. Our results indicate the presence of a symmetry-protected topological order coexisting with an antiferromagnetic order in this material.
\end{abstract}

\vskip2pc

\maketitle
The quest for novel phases of matter, particularly phases of matter beyond Landau's paradigm~\cite{Landau37:11,Ginzburg50:20} known as topological phases of matter after the discovery of the fractional quantum Hall effect~\cite{Tsui82:48,Laughlin83:50}, is now a profound and influential subject in condensed matter physics. Some of them have also been proposed as platforms for topologically-protected quantum computing~\cite{Kitaev03:303,Nayak08:80}. One important concept in this research field is that of a symmetry-protected topological phase of matter. Such phases are not restricted to nearly free fermion systems such as topological insulators~\cite{Kane05:95-1,Kane05:95-2,Bernevig06:96,Hasan10:82} and superconductors~\cite{Schnyder08:78,Sato10:80}, but can also be found in strongly interacting systems. The antiferromagnetic (AFM) \emph{S}=1 Heisenberg chain~\cite{Haldane83:93,Haldane83:50} turns out to be the first and simplest example of a symmetry-protected topological phase~\cite{Chen13:87,Wen17:89}. It is noteworthy that in this system an \emph{S}=1 spin can be formed out of two \emph{S}=1/2 spins, as has been shown rigorously for the Afﬂeck-Kennedy-Lieb-Tasaki model~\cite{Affleck87:59,Affleck88:115}. Thus, \emph{S}=1/2 two-leg ladder systems emerge as a realizable platform for the Haldane phase~\cite{Hida92:45,White96:53,Kim00:62}. Experimental evidence for the Haldane phase has been found in an AFM \emph{S}=1/2 Heisenberg ladder formed by cold atoms~\cite{Sompet22:664} but is still lacking in solid state materials.

\begin{figure}
\includegraphics[width=10cm,bbllx=30,bblly=150,bburx=580,bbury=650,angle=-90,clip=]{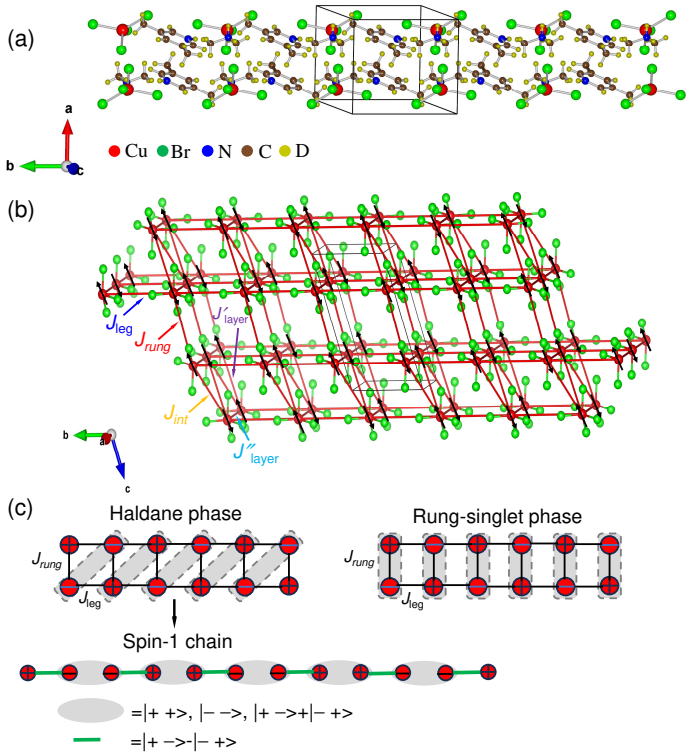}
\caption{Crystal structure and magnetic interactions in C$_9$H$_{18}$N$_2$CuBr$_4$. (a) Crystal structure of deuterated C$_9$H$_{18}$N$_2$CuBr$_4$ projected along the crystallographic \emph{c}-axis to show the stacking of discrete DMA$^+$ (C$_2$D$_{8}$N) and 35DMP$^+$ (C$_7$D$_{10}$N) cations. Outlined is a unit cell of the nuclear structure. (b) The three-dimensional exchange interactions between Cu$^{2+}$ ions including the intra-ladder couplings $J_{\rm rung}$ and $J_{\rm leg}$, the inter-ladder coupling $J_{\rm int}$, the inter-layer couplings $J^\prime_{\rm layer}$ (frustrating) and $J^{\prime\prime}_{\rm layer}$. The organic cations are not shown for the sake of simplicity. Black arrows indicate the directions of the spins in the N$\rm \acute{e}$el-ordered phase. (c) (Top) Schematic diagram showing two possible mappings of the AFM \emph{S}=1/2 Heisenberg two-leg ladder, into the Haldane and the rung-singlet phases. (Bottom) Representation of the valence bond ground state of the AFM spin-1 chain. Gray ovals and short green horizontal lines denote \emph{S}=1 spins and singlet pairs of auxiliary \emph{S}=1/2 spins, respectively.} \label{fig1}
\label{fig1}
\end{figure}

The \emph{S}=1/2 magnetic insulator C$_9$H$_{18}$N$_2$CuBr$_4$, DLCB for short, is promising in this regard. As shown in the crystal structure in Fig.~\ref{fig1}(a), CuBr$_4^{2-}$ radicals form a two-leg spin ladder with the ladder direction extending along the crystallographic \emph{b}-axis~\cite{Awwadi08:47}. An AFM ordered phase of DLCB below the transition temperature $T_{\rm N}$=2.0(1) K was observed in the specific heat and neutron diffraction measurements~\cite{Hong14:89}. The spin structure is collinear with staggered moments pointing alternately along an easy axis ($\equiv$$\hat{z}$), i.e., the \textbf{c}$^\ast$-axis in the reciprocal space, with a reduced ordered moment size of 0.39(5) $\mu_{\rm B}$. The polarized neutron study~\cite{Hong17:13} confirms that the gapped triplet ($S$=1 and $S_z$=0, $\pm1$) excitation energy splits into a gapped doublet ($S$=1 and $S_z$=$\pm1$) and a gapped nondegenerate "singlet" ($S$=1 and $S_z$=0), which are interpreted as the transverse mode (TM) and the longitudinal mode (LM) reflecting spin fluctuations perpendicular and parallel to the easy axis, respectively. Importantly, analysis of the spin Hamiltonian suggests that DLCB is close to the quantum critical point at ambient pressure and zero field~\cite{Hong14:89,Hong17:13,Ying19:122}. Thus, its magnetic properties could be extraordinarily responsive to an external stimulus such as hydrostatic pressure. Recent work~\cite{Hong22:13} reveals that the N$\rm \acute{e}$el-ordered phase breaks down beyond a critical pressure of $P_{\rm c}$$\sim$1.0 GPa. Estimates of the critical exponents, along with the broad spectral linewidth observed near the phase transition, suggest that the emergence of fractionalized excitations through the pressure-induced quantum phase transition (QPT) may be realized in DLCB at the critical pressure~\cite{Hong22:13}. Based on the crystal structure, we propose a minimal spin Hamiltonian of a three-dimensional (3D) frustrated interaction network of quantum spins as shown in Fig.~\ref{fig1}(b), where the inter-ladder coupling $J_{\rm int}$, the inter-layer couplings $J^\prime_{\rm layer}$ (frustrating) and $J^{\prime\prime}_{\rm layer}$ form triangular spin arrangements. However, due to the limited instrumental resolution, the previous work~\cite{Hong22:13} could not firmly establish if it is a continuous (second-order) or a pseudo-universal weakly first-order phase transition. The true nature of this QPT remains to be settled definitively. As detailed in the Supplementary Material~\cite{sm}, employing high-resolution neutron spectroscopy, we unambiguously identify a weakly first-order hydrostatic pressure-driven quantum phase transition, which arises from fluctuations enhanced by the frustrating interlayer coupling.

We verify that the AFM ordered phase of DLCB at ambient pressure is also unconventional, i.e., it cannot be completely characterized by a local order parameter. We examine its bulk thermodynamic properties and temperature dependence of spin dynamics, and find no evidence of either a conventional spin-flop transition in the magnetization curve or conventional magnons in neutron scattering spectra. With support from the adiabatic transformation between the \emph{S}=1 chain and the AFM \emph{S}=1/2 ladder system, we conclude that the spin gap in DLCB is not due to the spin anisotropy but related to that of the Haldane gap of \emph{S}=1 Heisenberg chains in a topologically nontrivial phase. This observation supports that the ground state of DLCB is best described as a quantum superposition of a N$\rm \acute{e}$el-ordered phase and a Haldane phase.

\begin{figure}[h]
\includegraphics[width=9.5cm,bbllx=102,bblly=295,bburx=502,bbury=530,angle=-90,clip=]{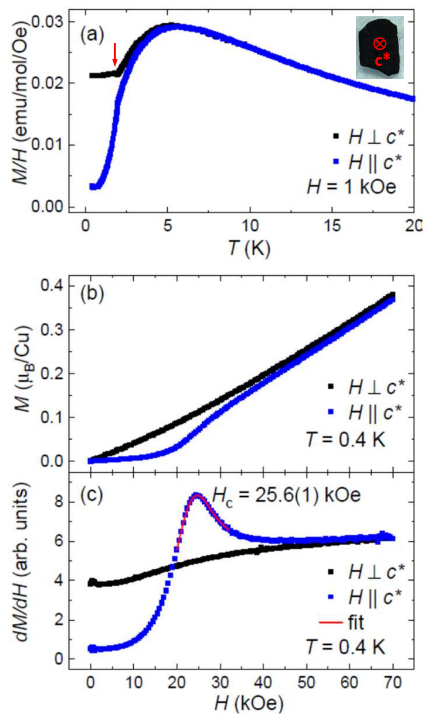}
\caption{(a) Magnetic susceptibility of DLCB measured in a 1 kOe magnetic field applied perpendicular and parallel to the easy axis (the \textbf{c}$^\ast$-axis). The red arrow indicates the transition at 2.0 K. Inset: A plate-shaped single-crystal sample of DLCB. The easy axis was confirmed by the neutron diffraction to be out of the flat face. (b) Magnetization \emph{M} and (c) the field derivative, d\emph{M}/d\emph{H}, at \emph{T}=0.4 K for \emph{H}$\perp$c$^\ast$ and \emph{H}$\parallel$c$^\ast$. The critical field $H_{\rm c}$ is determined at the maximum of d\emph{M}/d\emph{H} for \emph{H}$\parallel$c$^\ast$. The data were acquired while ramping the field in both directions, and no hysteresis was observed.}
\label{fig2}
\end{figure}
First, we show the susceptibility data in Fig.~\ref{fig2}(a) with a small magnetic field applied parallel and perpendicular to the \textbf{c}$^\ast$-axis. There is a rounded maximum near 5.7 K, which can be attributed to a triplet excitation~\cite{Troyer94:50}. Anisotropic behavior starts to develop below this broad maximum: $\chi_{\perp}$ has a kink at $T_{\rm N}$=2.0 K and roughly keeps the value at $T_{\rm N}$ down to 0.4 K whereas $\chi_{\parallel}$ decreases quickly below $T_{\rm N}$ due to the presence of a spin gap. In case of a conventional two-sublattice antiferromagnet with a weak Ising anisotropy, a spin-flop transition, indicated by a vertical jump in magnetization, is expected when the field is parallel to the easy axis. However, there is no such a characteristic jump nor a hysteresis in the magnetization of DLCB for \emph{H}$\parallel$c$^\ast$ in Fig.~\ref{fig2}(b). The field-derivative curve of magnetization d\emph{M}/d\emph{H} at \emph{H}$\parallel$c$^\ast$ in Fig.~\ref{fig2}(c) exhibits a definite anomaly at a critical field $H_{\rm c}$ of 25.6 kOe. It corresponds to a spin gap of $g\mu_{\rm B}H_{\rm c}\simeq$0.32 meV, where \emph{g}=2.15~\cite{Hong17:13} is the Land$\acute{\rm e}$ \emph{g} factor and $\mu_{\rm B}$ is the Bohr magneton. As there is no evidence of a conventional spin-flop transition in the magnetization for \emph{H}$\parallel$c$^\ast$ and the magnetization curve instead resembles those of the \emph{S}=1/2 two-leg ladder~\cite{Schmidiger12:108} and the \emph{S}=1 Haldane chain~\cite{Ajiro89:63} (see the Supplementary Material~\cite{sm} for details), indicating that the observed spin gap is not due to the spin anisotropy but to the separation between a singlet ground state and a triplet excited state.

\begin{figure*}[t]
\includegraphics[width=8cm,bbllx=133,bblly=60,bburx=453,bbury=740,angle=-90,clip=]{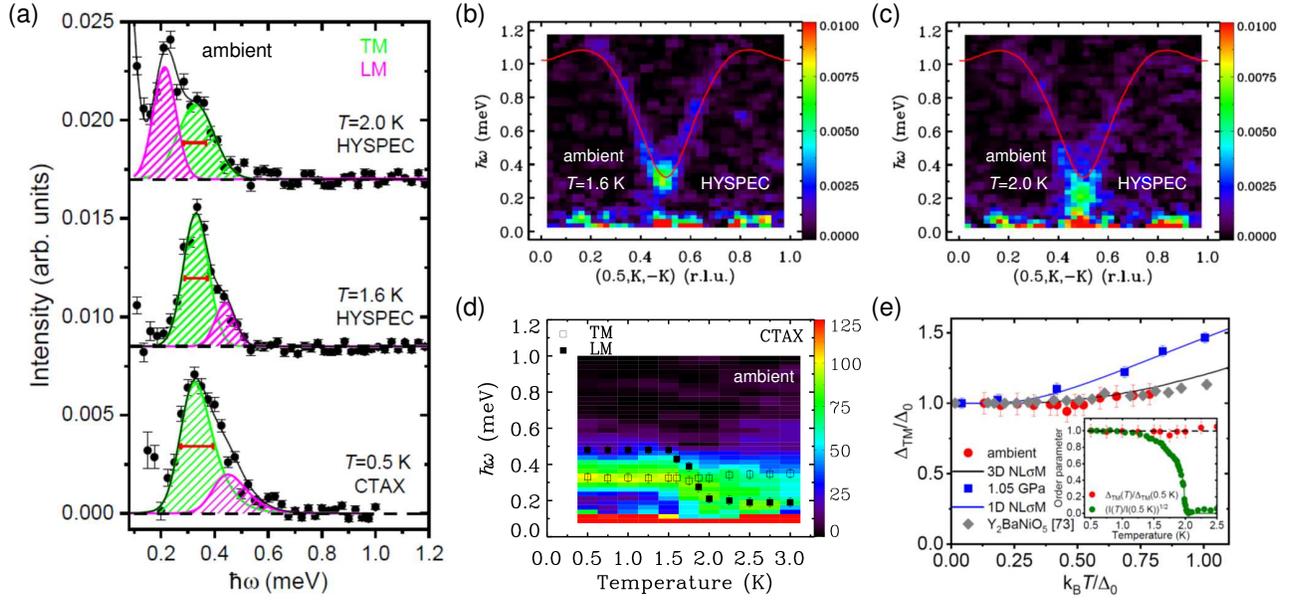}
\caption{(a) The representative ambient pressure background-subtracted transferred energy scans at the AFM wavevector \textbf{q}=(0.5,0.5,-0.5) at different temperatures. For clarity, the data are shifted upwards. The intensity at \emph{T}=0.5 K is normalized between two different instruments. The green and magenta shaded areas represent contributions from the TM and LM, respectively. The black solid lines are their sum. The red horizontal bars represent the instrumental resolution. The black dashed lines are guides to the eye. False-color maps of the magnetic excitation spectra as a function of wavevector and energy transfer at ambient pressure at (b) \emph{T}=1.6 K and (c) \emph{T}=2.0 K. The red lines are the linear spin wave theory calculations of the acoustic TM at \emph{T}=0~\cite{Toth15:27}. (d) False-color map of inelastic intensity at \textbf{q}=(0.5,0.5,-0.5) as a function of temperature and energy transfer at ambient pressure. Open and filled symbols are the obtained temperature dependence of the gap energy for TM and LM, respectively. (e) The reduced gap energy plotted against the reduced temperature at ambient pressure (red circle) and \emph{P}=1.05 GPa (blue square) and for Y$_2$BaNiO$_5$ (gray diamond)~\cite{Xu07:317}. The black and blue lines are calculations for the 3D and 1D nonlinear-$\sigma$ models, respectively. Inset: Normalized energy gap of the TM and the square root of the magnetic scattering intensity at \textbf{q}=(0.5,0.5,-0.5) vs. temperature at ambient pressure. The black dashed and olive solid lines are guides to the eye.} \label{fig3}
\end{figure*}

To gain more insight into the origin of the spin gap, we performed single-crystal inelastic neutron scattering at ambient pressure to investigate the effect of temperature on the spin dynamics of DLCB. Figure~\ref{fig3}(a) shows the representative background-subtracted energy scans at the AFM wavevector \textbf{q}=(0.5,0.5,-0.5) for several temperatures. At \emph{T}=0.5 K, the best fit yields the gap energies of the resolution-limited TM and LM as $\Delta_{\rm TM}$=0.33(3) meV and $\Delta_{\rm LM}$=0.46(3) meV. Their values are consistent with the previously reported values~\cite{Hong17:13}. While warming up towards $T_{\rm N}$, the scattering intensity of LM increases and $\Delta_{\rm LM}$ becomes softened. Its value is reduced to 0.44(3) meV and 0.21(3) meV at \emph{T}=1.6 K and 2.0 K, respectively. The spectrum at 2.0 K also exhibits quasi-elastic scattering, which is due to the fact that the magnetic Bragg peak disappears at $T_{\rm N}$ and its spectral weight spreads over to a small-energy transfer regime to satisfy the total moment sum rule. We find that the spectral lineshape of the TM gap becomes broad upon approaching $T_{\rm N}$ in contrast with the LM that is limited by the instrumental resolution. The intrinsic linewidth of $\Delta_{\rm TM}$ comes with a FWHM of 0.04(3) meV and 0.10(3) meV at \emph{T}=1.6 K and 2.0 K, respectively. Figures~\ref{fig3}(b-c) show false-colour maps of the magnetic excitation spectra measured at \emph{T}=1.6 and 2.0 K. The red lines are the linear spin wave theory calculations using SPINW~\cite{Toth15:27}. Apparently, the entire branch of the TM excitation is affected by such thermal broadening.

In the broken-symmetry phase (ordered phase), the low-energy collective excitations can be categorized by the phase and amplitude oscillations of the local order parameter. The phase oscillations are the gapless Goldstone modes~\cite{Goldstone62:127} or the TMs as spin waves (\emph{S}=1 magnons) of the broken continuous symmetry and the amplitude oscillation is the LM~\cite{Affleck92:46,Normand97:56,Lake20:85,Ruegg08:100,Merchant14:10,Pekker15:6,Hong17:08,Jain17:13,Zhu19:01,Hayashida19:5,Do21:12,Do23:8}. In case of a small Ising-type anisotropy, symmetry is broken not continuously but explicitly so that the TMs acquire an energy gap as gapped magnons. Note that, in a conventional 3D ordered phase, evolution of the TM with temperature below $T_{\rm N}$ in the two-sublattice antiferromagnet with a weak easy axis anisotropy is directly related to the AFM order parameter as $\Delta_{\rm TM}(T)$$\propto$$<$\emph{S}$>$ and hence $\Delta_{\rm TM}$ is expected to collapse at the transition temperature~\cite{Prozorova69:28,Harris01:226}. Such behavior was confirmed experimentally, e.g., in the 3D antiferromagnet MnF$_2$~\cite{Hagiwara96:8,Yamani10:88}, the quasi-two-dimensional square lattice antiferromagnet Sr$_2$Cu$_3$O$_4$Cl$_2$~\cite{Kim99:83,Kim01:64}, and the quasi-one-dimensional chain antiferromagnet Sr$_2$CuO$_3$~\cite{Sergeicheva17:95}. The determined temperature dependence of the energy gap $\Delta_{\rm TM}$, shown in Fig.~\ref{fig3}(d), however, suggests that $\Delta_{\rm TM}$ is seen to maintain its \emph{T}=0.5 K value and lifts off slowly right above the transition. The inset of Fig.~\ref{fig3}(e) shows $\Delta_{\rm TM}(T)/\Delta_{\rm TM}$(0.5 K), along with the square root of magnetic Bragg peak \emph{I(T)}/\emph{I}(0.5 K) at \textbf{q}=(0.5,0.5,-0.5), which also probes the AFM order parameter. From the perspective of spin waves for a collinear antiferromagnet, the marked contrast between $\Delta_{\rm TM}(T)/\Delta_{\rm TM}$(0.5 K) and \emph{I(T)}/\emph{I}(0.5 K) provides further evidence that the TM gap in DLCB is not owing to the spin anisotropy.

\begin{figure}[t]
\includegraphics[width=6.5cm,bbllx=145,bblly=220,bburx=460,bbury=610,angle=-90,clip=]{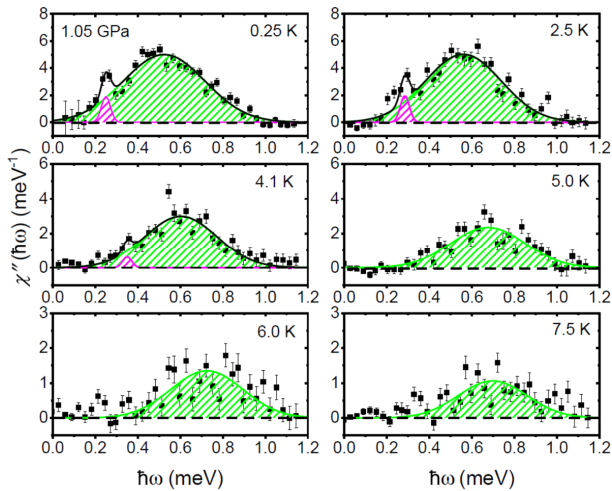}
\caption{Normalized imaginary part of the dynamical spin susceptibility at 1.05 GPa at \emph{T}=0.25 K, 2.5 K, 4.1 K, 5.0 K, 6.0 K, and 7.5 K. The data were collected at CNCS and normalized to absolute units by the total moment sum rule (see the Supplementary Material~\cite{sm} for details). The green and magenta shaded areas represent contributions from the TM and LM, respectively. Note that the LM is traceable up to 4.1 K. The black solid lines at \emph{T}=0.25 K, 2.5 K, and 4.1 K are their sum. The black dashed lines are guides to the eye.}\label{fig4}
\end{figure}

In Fig.~\ref{fig3}(e), we plot $\Delta_{\rm TM}$ with respect to its value $\Delta_0$ at the lowest temperature, $\delta$=$\Delta_{\rm TM}(T)/\Delta_0$, as a function of the reduced temperature $\tau$=$k_{\rm B}T/\Delta_0$, $k_{\rm B}$ being the Boltzman constant. At $k_{\rm B}T$$<$$\Delta$, a universal field theory with a single energy scale $\Delta$ can be derived by a large-\emph{S} mapping of the Heisenberg Hamiltonian onto the \emph{O}(3) nonlinear $\sigma$ model~\cite{Senechal93:47,Damle98:57}. It offers a good description of the finite-\emph{T} renormalization of excitations in the quantum disordered phase of gapped antiferromagnets~\cite{Zheludev08:100}, where a spin gap separates the singlet ground state from the triplet excitation. The scaled gap energy with temperature at 1.05 GPa (blue square in Fig.~\ref{fig3}(e), the relevant neutron data are plotted in Fig.~\ref{fig4}) in the quantum disordered phase indeed shows a good agreement with the theoretical calculation for the one-dimensional (1D) nonlinear $\sigma$ model~\cite{Damle98:57}. This is consistent with the fact that the exchange coupling ratio of $J_{\rm int}$/$J_{\rm leg}$ gets reduced with pressure~\cite{Hong22:13} thus the system becomes quasi-1D, interladder couplings becoming unimportant, at the criticality. As a result, the remarkable agreement between the experimental result at ambient pressure (red circle) and the calculation for the 3D nonlinear-$\sigma$ model~\cite{Senechal93:47} indicates that a singlet state with a spin gap coexists with the N$\rm \acute{e}$el-ordered phase in DLCB. In the subsequent discussion, we will examine this counterintuitive result.

Since there cannot be any single-ion anisotropy for the \emph{S}=1/2 Cu$^{2+}$ ions, an Ising-type anisotropy is necessary to account for the small split between the TMs and LM in DLCB. However, it can be ruled out that the spin gap in DLCB comes from the disordered \emph{S}=1/2 ladder with Ising-type anisotropy, where $\Delta_{\rm LM}$ is smaller than $\Delta_{\rm TM}$~\cite{Blosser19:100}, contrary to our finding for DLCB. In the limit of $J_{\rm rung}$$>>$$J_{\rm leg}$, the ground state of the AFM \emph{S}=1/2 ladder can be obtained perturbatively starting from a product of rung singlets. However, when $J_{\rm rung}$$\approx$$J_{\rm leg}$, there is no small parameter to conduct a perturbative analysis~\cite{White94:73,Dagotto96:271} and thus it is necessary to consider also other dimer configurations. As shown in Fig.~\ref{fig1}(c), the AFM \emph{S}=1/2 ladder can be adiabatically transformed either to the composite spin representation of the \emph{S}=1 chain by pairing two spins diagonally or to a conventional rung-singlet phase by pairing two spins vertically, along the rung direction. In the case of the Haldane phase, the \emph{S}=1 spin on each lattice site (the gray oval) can be decomposed into two \emph{S}=1/2 auxiliary spins, which form singlets with nearest neighbors. At the two ends of an open chain, there are two unpaired \emph{S}=1/2's behaving like free \emph{S}=1/2 moments, forming the zero-energy modes. We can make a description in terms of total spin per \emph{i}th unit cell, \textbf{S}$_i$=\textbf{S}$_{i,1}$+\textbf{S}$_{i,2}$, where the indices (1, 2) indicate the two \emph{S}=1/2 spins in the same unit cell, making an effective \emph{S}=1 spin. Numerical results~\cite{White96:53,Sompet22:664} indicate that even without a diagonal coupling, the probability to find a diagonally situated pair of spins in a triplet state is $\sim$96$\%$ for $J_{\rm rung}$/$J_{\rm leg}$=1 and has a maximum at $J_{\rm rung}$/$J_{\rm leg}$$\simeq$1.3, close to 1.1 in DLCB~\cite{Hong14:89}. The ground state in this case is a valence-bond solid~\cite{Affleck87:59,Affleck88:115}. The excitations on top of this ground state involve flipping the singlets into triplets and cost finite energy. In light of this, we conclude that the gap in DLCB is a Haldane gap, initially proposed by Haldane to describe the low-energy dynamics of the \emph{S}=1 AFM Heisenberg chain, whose ground state is a symmetry-protected topological phase. Indeed, a similar temperature dependence of the gap energy (grey diamond in Fig.~\ref{fig3}(e)) was reported for the \emph{S}=1 chain compound Y$_2$BaNiO$_5$~\cite{Xu07:317}. In the case of the rung-singlet phase, numerical results~\cite{Sompet22:664} indicate that the probability to find a singlet on the rung of the ladder is $\sim$60$\%$ for $J_{\rm rung}$/$J_{\rm leg}$=1.1. Its ground state is quantum disordered and there is a spin gap for all nonzero value of $J_{\rm rung}$. The system can be driven, through a quantum phase transition, into a magnetically ordered gapless phase by a finite interladder coupling $J_{\rm int}$~\cite{Capriotti02:65}. Our previous work~\cite{Hong14:89} shows that $J_{\rm int}$ of DLCB is very near the critical value required to drive the system to a N$\rm \acute{e}$el-ordered phase. Altogether, the ground state of DLCB is best described as a quantum superposition of a Haldane phase and a N$\rm \acute{e}$el-ordered phase, which resembles the quantum state of a qubit in quantum computing as a linear superposition of two orthonormal basis states. It is also essential to note that the LM, i.e., amplitude fluctuations of the order parameter, is long-lived against thermal fluctuations and does not become gapless at the phase transition, as shown in Fig.~\ref{fig3}(d), unlike at a classical phase transition~\cite{Merchant14:10}. The finite energy gap of $\Delta_{\rm LM}$=0.21(3) meV at $T_{\rm N}$ indicates a weakly first-order thermal phase transition. In fact, the presence of an order parameter coupling to some fluctuating field can convert a continuous phase transition to a weakly first-order transition~\cite{Halperin74:32,She10:82}.

In summary, we have performed bulk thermodynamics and high-resolution neutron scattering measurements on the \emph{S}=1/2 two-leg ladder antiferromagnet DLCB. Our most significant finding is that the ground state of DLCB is best described as a quantum superposition of a N$\rm \acute{e}$el-ordered phase and a Haldane phase. Our experimental and theoretical results establish DLCB as an ideal experimental platform for observing a new type of symmetry-protected topological phase of matter.

\begin{acknowledgments}
We gratefully acknowledge the helpful discussions with Lakshmi Bhaskaran, Yongqiang Cheng, Satoshi Okamoto, Stefan Wessel, Cenke Xu, Tao Ying, Igor Zaliznyak, and Sergei Zvyagin. We also thank Matthew Collins, Michael Cox, Saad Elorfi, Cory Fletcher, Melissa Graves-Brook, Kostya Nasyedkin, Christopher Redmon, Christopher Schmitt, Todd Sherline, and Tyler White for the technical support in neutron scattering experiments. A portion of this research used resources at the High Flux Isotope Reactor and Spallation Neutron Source, a DOE Office of Science User Facility operated by the Oak Ridge National Laboratory (ORNL). Magnetization measurements (A.F.M.) were supported by the U. S. Department of Energy, Office of Science, Basic Energy Sciences, Materials Sciences and Engineering Division. Work at Michigan State University is support by the U.S. Department of Energy, Office of Science, Office of Basic Energy Sciences, Materials Sciences and Engineering Division under DE-SC0019259. MD acknowledges funding from the Spanish Ministerio de Ciencia e Innovaci$\rm \acute{o}n$ (Project PID2020-117803GB-I00) and the Generalitat de Catalunya (project grant 2021SGR00354).
\end{acknowledgments}


\noindent
\section{\textbf{Supplementary Material}}

\beginsupplement

\section{Experimental Methods}
Deuterated single crystals were grown using a solution method [1]. An aqueous solution containing a 1:1:1 ratio of deuterated (DMA)Br, (35DMP)Br, where DMA$^+$ is the dimethylammonium cation and 35DMP$^+$ is the 3,5-dimethylpyridinium cation, and the corresponding copper(II) halide salt was allowed to evaporate for several weeks; a few drops of DBr were added to the solution to avoid hydrolysis of the Cu(II) ion.

Magnetization measurements were performed in a Quantum Design MPMS3 from 300 to 0.4 K with the magnetic field applied either parallel or perpendicular to the easy axis. The easy axis, i.e., the \textbf{c}$^\ast$-axis, was confirmed by the neutron diffraction to be out of the flat face of the single-crystal sample. Possible misalignment of the magnetic field relative to the easy axis was within 2$^\circ$. The data below 1.8 K were collected using the helium-3 insert. In isothermal measurements at \emph{T}=0.4 K, no hysteresis was observed between data collected upon increasing and decreasing the magnetic field. The data taken with the helium-3 insert were scaled at 10 K to those taken without it, by using the dataset for \emph{H}$\parallel$c$^\ast$.

Single-crystal neutron diffraction measurements under pressure were carried out on a cold neutron triple-axis spectrometer (CTAX) with both the incident and final neutron energies fixed at 4.5 meV at High Flux Isotope Reactor (HFIR), Oak Ridge National Laboratory (ORNL). A cooled Be filter was placed after the sample to eliminate higher-order beam contamination. A helium-3 cryostat insert was used to achieve the base temperature of 0.25 K. Neutron scattering data were also collected using a cold neutron chopper spectrometer (CNCS)~[2] at the Spallation Neutron Source (SNS), ORNL. The scattering intensity was normalized to the number of incident protons per pulse and integrated out of the scattering plane direction by a narrow slice of $\pm$0.1 r.l.u. in order to analyze the experimental data taken in the scattering plane. The incident neutron energy was chosen as 2.07 meV and the energy resolution at the elastic line is 0.08 meV. The sample used for the pressure study consists of three co-aligned deuterated single crystals with a total mass of 0.25 g and a 1.5$^\circ$ mosaic spread. The sample was loaded inside a CuBe piston-cylinder-type pressure cell with maximum allowable hydrostatic pressure $\sim$1.3 GPa. Fluorinert FC-770 was used as pressure-transmitting medium to achieve good hydrostaticity. The desired hydrostatic pressure was applied by a hydraulic press. Change of volume with pressure in NaCl was used for pressure calibration with an accuracy of 0.1 GPa. Inelastic neutron scattering data at ambient pressure were collected on CTAX, HFIR and the hybrid spectrometer (HYSPEC)~[3], SNS using a standard helium-flow cryostat. The sample consists of two co-aligned deuterated single crystals with a total mass of 2.5 g and a 1.0$^\circ$ mosaic spread. For the CTAX experiment, a PG(002) analyzer set to reflect neutrons with the final neutron energy 3.2 meV was used and the energy resolution at the elastic line is 0.12 meV. For the HYSPEC experiment, the incident neutron energy was chosen as 3.8 meV with a Fermi chopper frequency of 300 Hz, which gives the energy resolution of 0.1 meV at the elastic line. The sample was oriented in either the (H,H,L) or (H,K,$\bar{\rm K}$) scattering plane for the measurements under pressure or at ambient pressure. Neutron scattering intensity as shown in Figs.~2(c-d) of the main text was integrated along the H or L direction by $\pm$0.1 r.l.u or $\pm$0.2 r.l.u. Neutron scattering intensity as shown in Figs.~3(b-d) of the main text was integrated along the H or K direction by $\pm$0.1 r.l.u or $\pm$0.05 r.l.u. In all experiments, the background was determined at \emph{T}=15 K under the same instrumental configurations. The data sets collected at CNCS and HYSPEC were reduced and analyzed using the software package DAVE~[4].

\section{Thermodynamics Measurements}
\begin{figure}[h]
\includegraphics[width=4cm,bbllx=148,bblly=80,bburx=440,bbury=735,angle=-90,clip=]{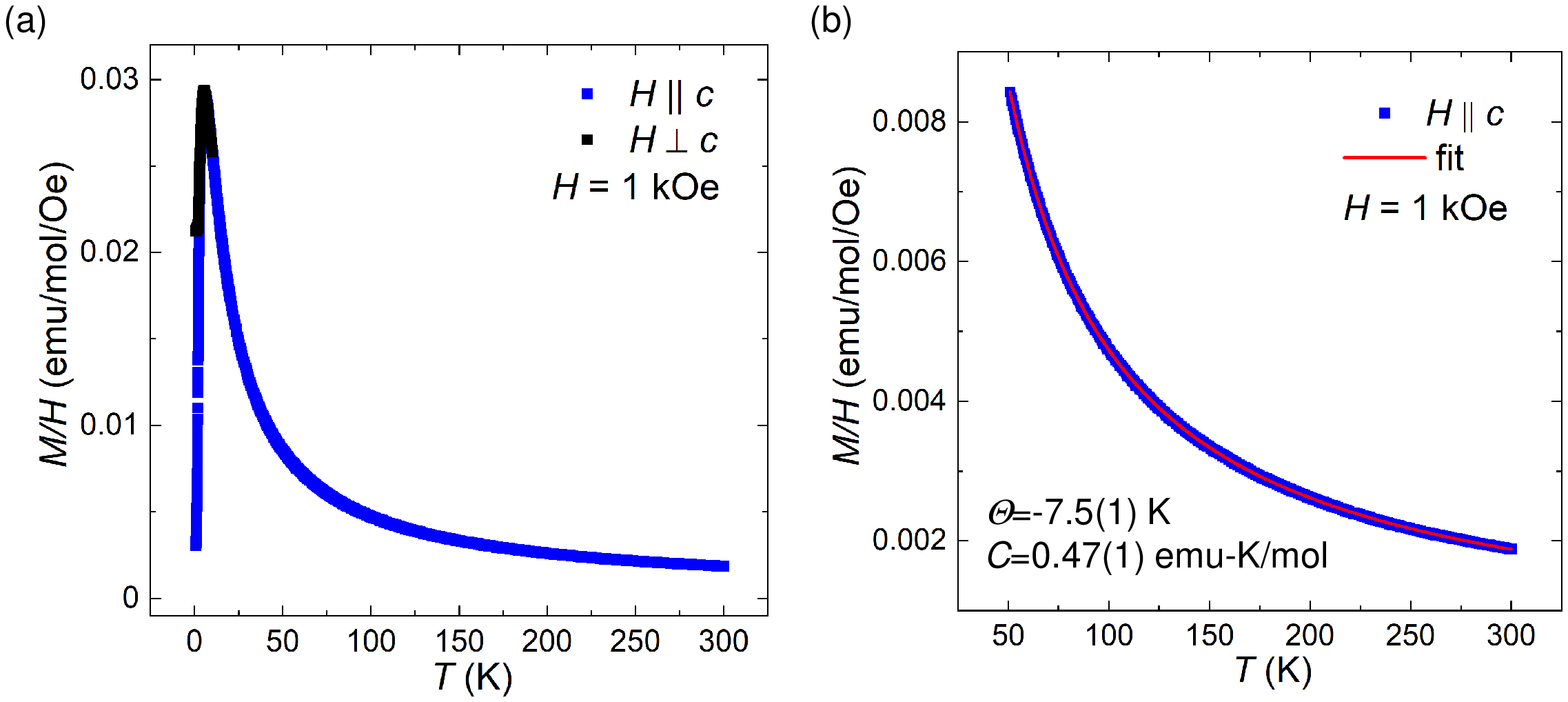}
\caption{(a) Overall magnetic susceptibility of DLCB measured in a 1 kOe magnetic field applied perpendicular and parallel to the easy axis i.e., the \textbf{c}$^\ast$-axis. (b) High-temperature data between 50 K and 300 K. The red line is a fit to the Curie‐Weiss law, yielding a Curie-Weiss temperature $\Theta$=-7.5(1) K.} \label{sus}
\end{figure}

The low temperature part of magnetic susceptibility data are plotted in Fig.~2(a) in the main text. Figure~\ref{sus}(a) shows the overall magnetic susceptibility with the extended temperature range up to 300 K. Within the temperature interval of 50 K to 300 K, the magnetic susceptibility in Fig.~\ref{sus}(b) is described by the Curie‐Weiss law as $\chi$(\emph{T})=\emph{C}/(\emph{T}+$\Theta$)+$\chi_0$, the best fit gives the Curie‐Weiss temperature $\Theta$=-7.5(1) K and the Curie-Weiss constant \emph{C}=0.47(1) emu-K/mol.

\begin{figure}
\includegraphics[width=5cm,bbllx=105,bblly=65,bburx=490,bbury=760,angle=-90,clip=]{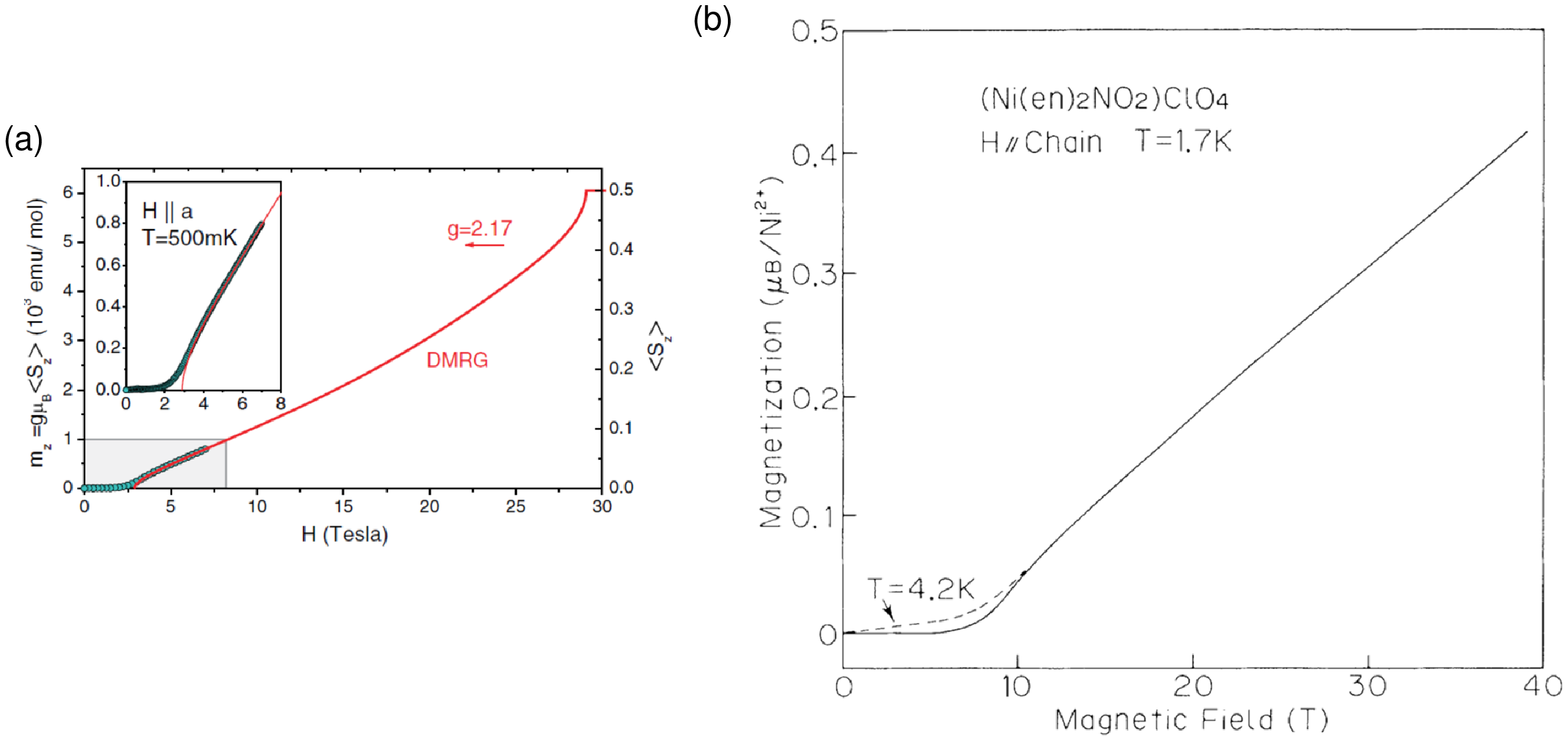}
\caption{(a) Magnetization of the antiferromagnetic \emph{S}=1/2 two-leg ladder material DIMPY for \emph{H}$\parallel$\emph{a}, adapted from Ref.~[5] and (b) Magnetization of the \emph{S}=1 chain material NENP in the magnetic field applied parallel to the chain direction, adapted from Ref.~[6].} \label{bulk}
\end{figure}

To compare with magnetization of DLCB for \emph{H}$\parallel$c$^\ast$ in Fig.~2(b) of the main text, we show in Fig.~\ref{bulk} the magnetization curves of the antiferromagnetic (AFM) \emph{S}=1/2 two-leg ladder material DIMPY~[5] and the \emph{S}=1 chain material NENP~[6].

\section{Single-Crystal Inelastic Neutron Measurements}
\begin{figure*}
\includegraphics[width=8cm,bbllx=135,bblly=80,bburx=472,bbury=760,angle=-90,clip=]{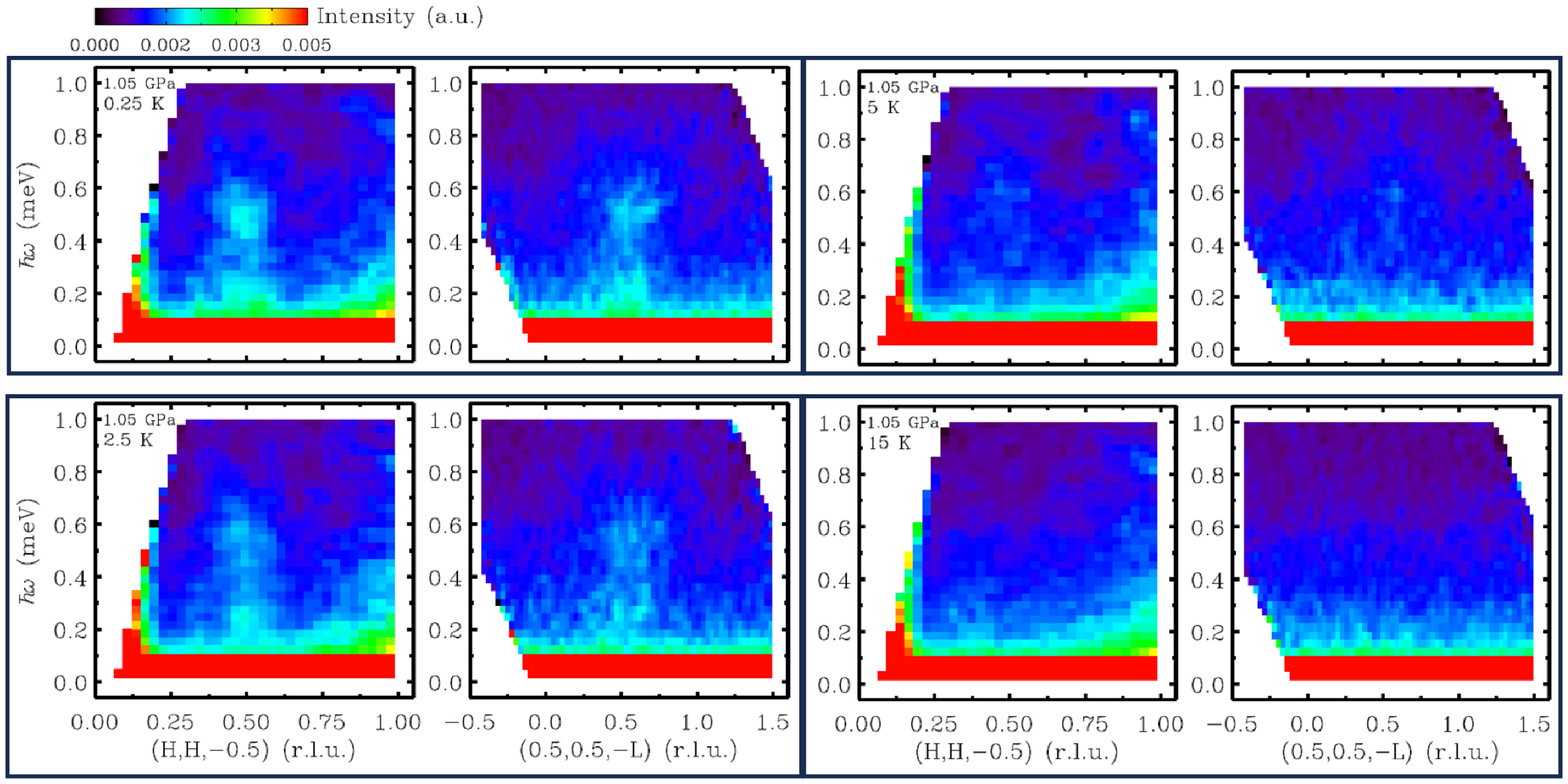}
\caption{False-color maps of the excitation spectra measured at CNCS as a function of energy and wavevector transfer along two high-symmetry directions (H,H,-0.5) and (0.5,0.5,-L) in the reciprocal space. The data were collected at \emph{P}=1.05 GPa and \emph{T}=0.25 K, 2.5 K, 5.0 K and 15.0 K.} \label{Spectra1}
\end{figure*}

\begin{figure*}
\includegraphics[width=8cm,bbllx=135,bblly=80,bburx=472,bbury=760,angle=-90,clip=]{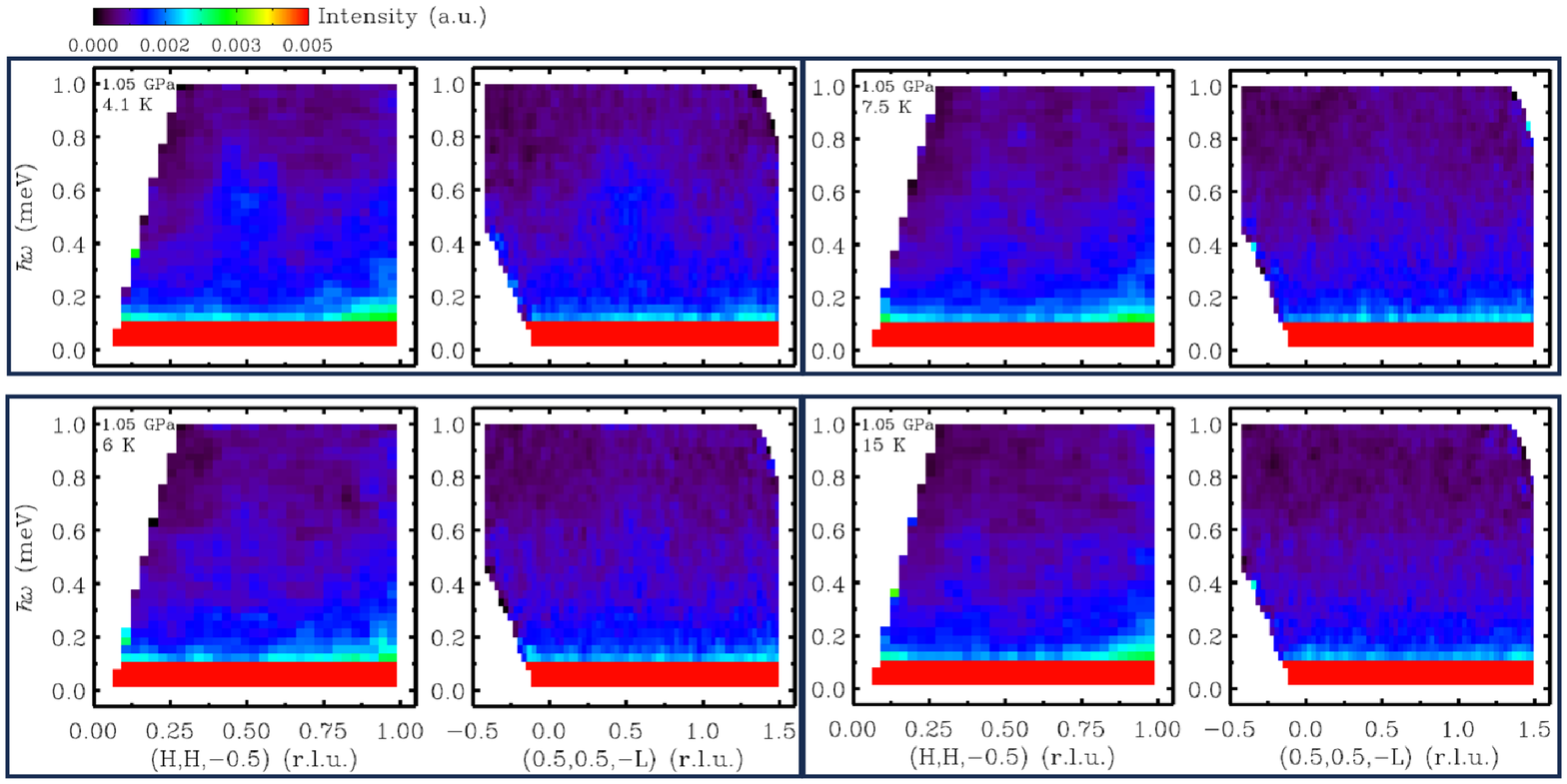}
\caption{False-color maps of the excitation spectra measured at CNCS in a separate experiment as a function of energy and wavevector transfer along two high-symmetry directions (H,H,-0.5) and (0.5,0.5,-L) in the reciprocal space. The data were collected at \emph{P}=1.05 GPa and \emph{T}=4.1 K, 6.0 K, 7.5 K and 15.0 K.} \label{Spectra2}
\end{figure*}

Figure~\ref{Spectra1} shows the neutron scattering intensity as a function of energy and wavevector transfer along two high-symmetry directions at \emph{P}=1.05 GPa and \emph{T}=0.25 K, 2.5 K, 5.0 K and 15 K. Neutron data were also collected using a standard helium-flow cryostat at the same hydrostatic pressure and \emph{T}=4.1 K, 6.0 K, 7.5 K and 15 K, as shown in Fig.~\ref{Spectra2}. The background at \emph{T}=15 K was measured for each experiment to account for the different sample-environment setups.


\begin{figure*}
\includegraphics[width=6.5cm,bbllx=160,bblly=100,bburx=410,bbury=715,angle=-90,clip=]{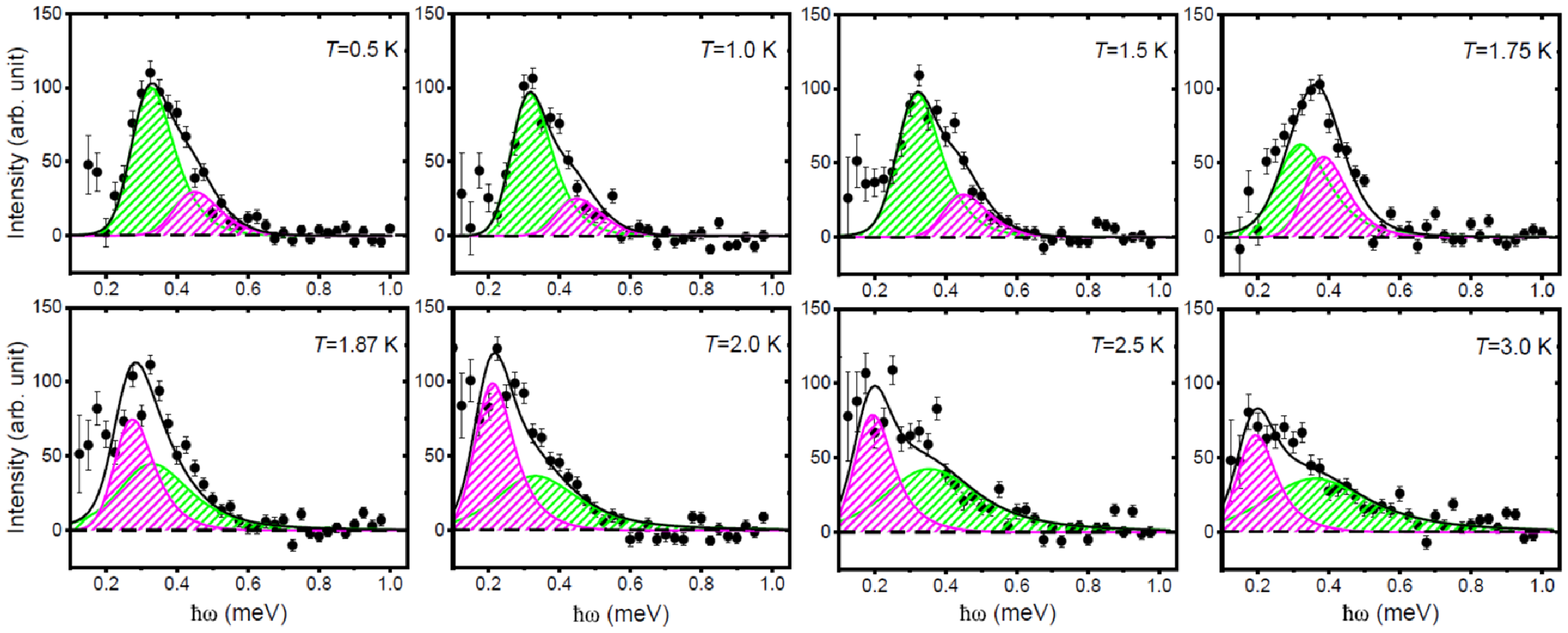}
\caption{The ambient pressure background-subtracted transferred energy scans at the antiferromagnetic wavevector \textbf{q}=(0.5,0.5,-0.5) at 0.5 K, 1.0 K, 1.5 K, 1.75 K, 1.87 K, 2.0 K, 2.5 K and 3.0 K, respectively. Neutron data were collected at CTAX. The green and magenta shaded areas represent contributions from the TM and LM, respectively. The black dashed lines are guides to the eye.} \label{Spectra3}
\end{figure*}

Figure~\ref{Spectra3} shows representative background-subtracted energy scans at the AFM wavevector \textbf{q}=(0.5,0.5,-0.5) for several temperatures to make the false-color map of Fig.~3(d) in the main text.

\vskip 1cm
\section{Estimates of the Inter-Layer Magnetic Interactions at Ambient Pressure}
\begin{figure*}
\includegraphics[width=8cm,bbllx=145,bblly=65,bburx=460,bbury=765,angle=-90,clip=]{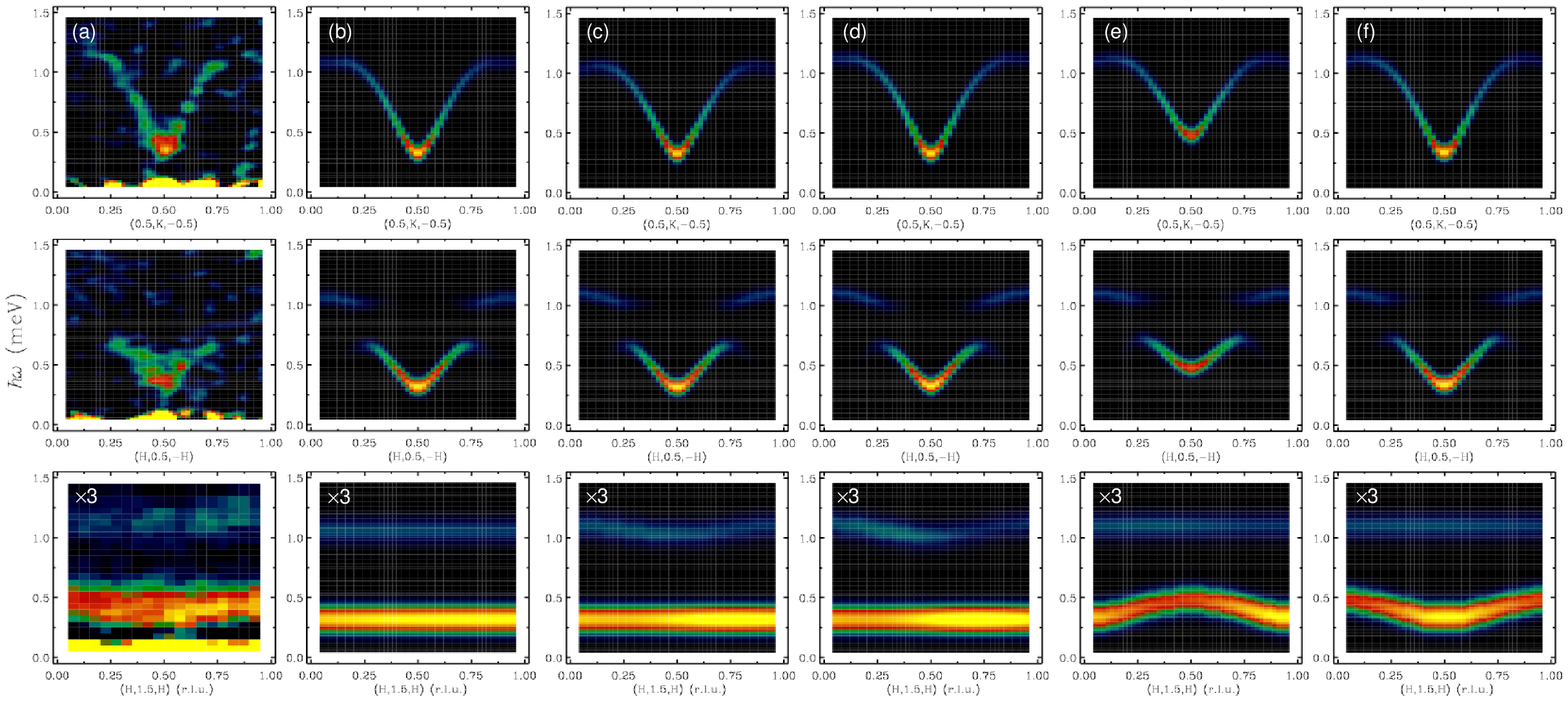}
\caption{False-colour maps of the magnetic excitation spectra along the ladder (top), between the ladder (middle), and between the two-dimensional layers (bottom) directions. Comparison between (a) the experimental data at ambient pressure at \emph{T}=1.45 K and the linear spin–wave theory calculations of the excitation spectra as a function of energy and wavevector transfer along three high-symmetric directions after convolving with the instrumental resolution function with (b) $J^\prime_{\rm layer}$=$J^{\prime\prime}_{\rm layer}$=0, (c) $J^\prime_{\rm layer}$=0.06 meV and $J^{\prime\prime}_{\rm layer}$=0.03 meV, (d) $J^\prime_{\rm layer}$=0.08 meV and $J^{\prime\prime}_{\rm layer}$=0.04 meV, (e) $J^\prime_{\rm layer}$=0.06 meV and $J^{\prime\prime}_{\rm layer}$=0, and (f) $J^\prime_{\rm layer}$=0 and $J^{\prime\prime}_{\rm layer}$=0.03 meV. To make the optical branch of the TM visible, the scattering intensity along the (H,1.5,H) direction was enlarged by a factor of 3. Note that the experimental data are adapted from Ref.~[7] and Supplemental Information of Ref.~[8].}\label{fig_appendix}
\end{figure*}
Based on the crystal structure of DLCB, we propose a minimal spin Hamiltonian of a 3D frustrated interaction network as shown in Fig. 1(b) of the main text. The magnetic interactions due to \emph{S}=1/2 Cu$^{2+}$ ions in DLCB are mediated by the superexchange across the diamagnetic bromide ions \emph{via} Cu-Br$\cdot\cdot\cdot$Br-Cu contacts. In terms of the Cu-Br$\cdot\cdot\cdot$Br bridging angle, as summed up by the Goodenough-Kanamori-Anderson rules, the larger the deviation from 180$^\circ$, the weaker is the antiferromagnetic superexchange and the coupling eventually becomes ferromagnetic for bridging angles close to 90$^\circ$. As their bridging angles are fairly larger than 90$^\circ$, the inter-layer exchange interactions $J^\prime_{\rm layer}$ and $J^{\prime\prime}_{\rm layer}$ are antiferromagnetic. While the coupling $J^{\prime\prime}_{\rm layer}$ along the \emph{a}-axis helps to align the ordered moments antiparallel to each other between adjacent layers in the N$\rm \acute{e}$el ordered phase, the coupling $J^\prime_{\rm layer}$ can lead to magnetic frustration.

Figure~\ref{fig_appendix}(a) shows false-colour maps of the magnetic excitation spectra along the ladder, between the ladders, and between the two-dimensional (2D) layers, respectively, measured at ambient pressure and \emph{T}=1.45 K. Note that the dispersion along the inter-layer direction is quite flat, indicating that the possible inter-layer couplings are expected to be weak. The best fit of the experimental data to the 2D unfrustrated spin interacting model yields $J_{\rm rung}$=0.70 meV, $J_{\rm leg}$=0.64 meV, $J_{\rm int}$=0.19 meV, and $\lambda$=0.95. The calculated dynamical structure factors using SPINW~[9] are plotted in Fig.~\ref{fig_appendix}(b). We further include $J^\prime_{\rm layer}$ and $J^{\prime\prime}_{\rm layer}$ in the 3D frustrated interaction network. In consideration of $J^\prime_{\rm layer}$ or $J^{\prime\prime}_{\rm layer}$ alone as shown in Figs.~\ref{fig_appendix}(e-f), the energy gap moves upwards and the AFM wavevector becomes maximum or minimum of the inter-layer dispersion. At $J^\prime_{\rm layer}$=2$J^{\prime\prime}_{\rm layer}$ as shown in Figs.~\ref{fig_appendix}(c-d), the dispersions along and between the ladder directions are not affected but modulates the intensity of the inter-layer dispersion. The values of $J^\prime_{\rm layer}$=0.06 meV and $J^{\prime\prime}_{\rm layer}$=0.03 meV provide a reasonable agreement with the experimental data.

\section{Hydrostatic Pressure-Induced Weakly First-Order Quantum Phase Transition}
\begin{figure}
\includegraphics[width=6cm,bbllx=100,bblly=60,bburx=500,bbury=615,angle=-90,clip=]{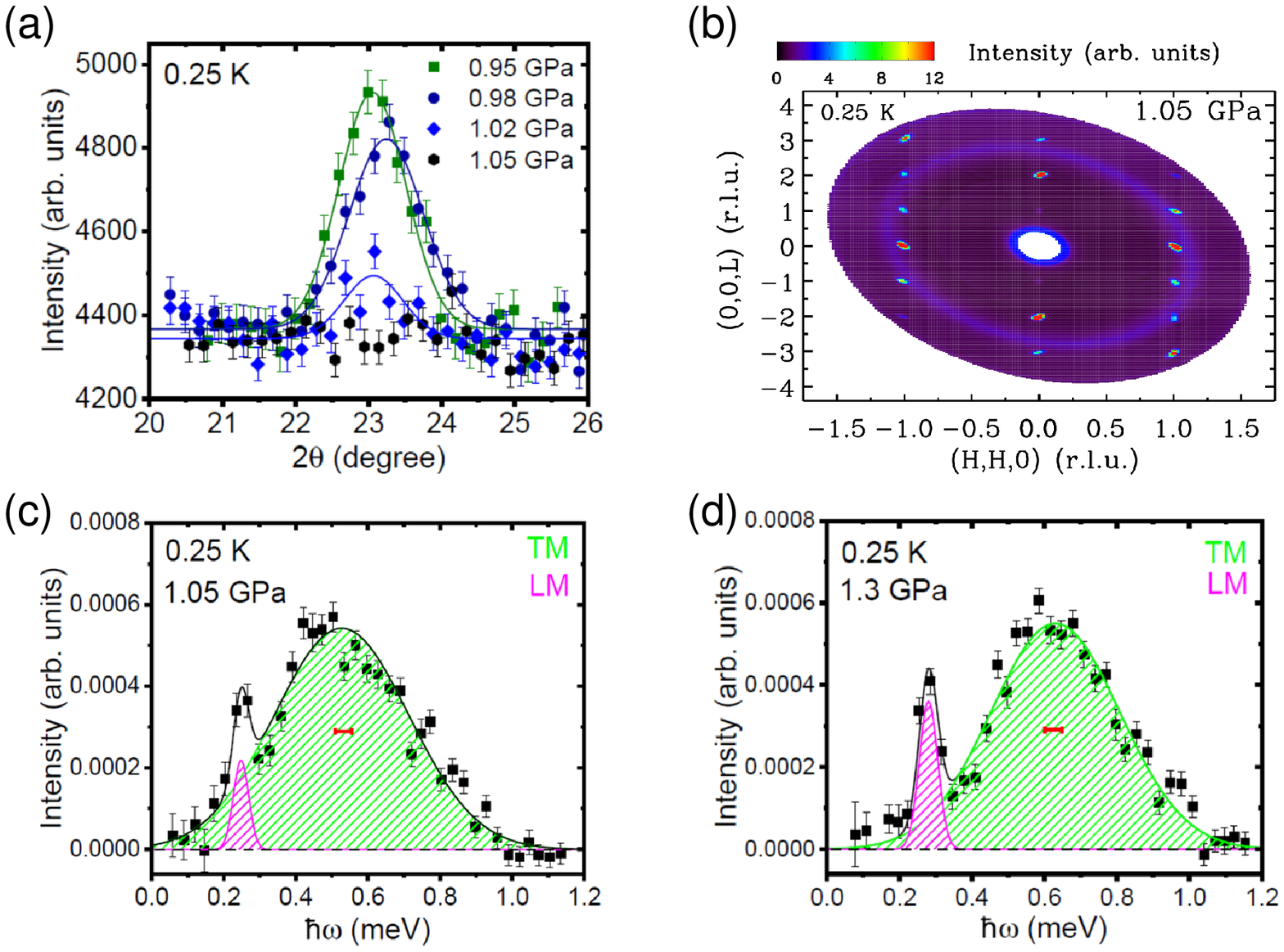}
\caption{Single-crystal neutron diffraction study under pressure. (a) The representative $\theta/2\theta$ scans measured at CTAX around the antiferromagnetic wavevector \textbf{q}=(0.5,0.5,-0.5) for 0.95, 0.98, 1.02, and 1.05 GPa, respectively. The solid lines are fits to the Gaussian profile. (b) Single-crystal neutron diffraction pattern measured at CNCS at \emph{T}=0.25 K and \emph{P}=1.05 GPa. The ring feature originates from the cytop glue. The local dynamic structure factors measured at CNCS for (c) 1.05 GPa and (d) 1.3 GPa, respectively, by integrating over momentum transfer along $0.4\leq \rm H\leq0.6$ and $-0.7\leq \rm L\leq-0.3$. The green and magenta shaded areas represent contributions from the TM and LM, respectively. The black solid lines are their sum. The black dashed lines are guides to the eye. The red horizontal bars represent the instrumental resolution. All data were collected at \emph{T}=0.25 K and the correction for the Cu$^{2+}$ magnetic form factor is included.}
\label{fig2}
\end{figure}
We first finely tune the ground state of DLCB with hydrostatic pressure to approach the quantum critical point (QCP) using the single-crystal neutron diffraction method. Figure~\ref{fig2}(a) shows the pressure-dependence of neutron diffraction $\theta$/2$\theta$ scans down to 0.25 K at the AFM wavevector \textbf{q}=(0.5,0.5,-0.5). The scattering intensity of the magnetic Bragg peak becomes continuously diminished as hydrostatic pressure increases and eventually disappears at \emph{P}$_{\rm c}$=1.05 GPa, in good agreement with the previous work~[10]. There is also no evidence of any incommensurate magnetic Bragg peak or diffuse scattering over a wide range of reciprocal space as shown in Fig.~\ref{fig2}(b) and the observed diffraction peaks are consistent with the triclinic space group P$\bar{1}$.

High-resolution inelastic neutron scattering was used in order to tell whether the low-energy magnetic excitation spectrum at \emph{P}$_{\rm c}$ is gapped or gapless. The background-subtracted energy scan at 1.05 GPa in Fig.~\ref{fig2}(c) clearly show two well-separated gapped modes including one sharp LM and another broad continuum-like TM excitation. The spectral lineshapes were modelled by superposition of two Gaussian distributions. The best fit yields the gap energies of the resolution-limited LM as $\Delta_{\rm LM}$=0.25(3) meV and a broad TM as $\Delta_{\rm TM}$=0.52(3) meV with a full width at half maximum (FWHM) of 0.44(3) meV (six times broader than the instrumental resolution), respectively. Such spectral broadening, indicative of an exotic pressure-induced gapped quantum paramagnetic phase, persists at least up to 1.3 GPa beyond the QCP, see Fig.~\ref{fig2}(d), where FWHM becomes 0.40(3) meV. Previous studies~[10] have shown that the observed continuum-like broad excitation of the TM in DLCB near the QPT cannot be reproduced by the dynamic structure factor calculated from the unfrustrated spin Hamiltonian~[11] or be attributed to spontaneous
quasiparticle decays~[12,13]. Consequently, the frustrating inter-layer coupling $J^\prime_{\rm layer}$ is necessary to be included in the original Hamiltonian (see the section III for details). On approaching the QCP, the weak $J^\prime_{\rm layer}$, that is deemed irrelevant initially, can become dominant. Frustration may enhance quantum fluctuations such that the transition is rendered weakly first-order~[14-16]. Indeed, as the correlation length is inversely proportional to $\Delta_{\rm LM}$~[17], a small energy gap of the LM at \emph{P}$_{\rm c}$ suggests a weakly first-order transition, where the correlation length is finite but much larger than the lattice spacings.

\section{Absolute Normalization of Magnetic Neutron Scattering Data}
In inelastic neutron scattering, the dynamic spin correlation function ${\cal S}({\rm \textbf{q}},\hbar\omega)$ describes the magnetic fluctuations in the sample, as a function of transferred momentum and energy. ${\cal S}({\rm \textbf{q}},\hbar\omega)$ satisfies the total moment sum rule when integrated over a Brillouin zone:
\begin{eqnarray}
\frac{\int_{-\infty}^{+\infty}\int_{\rm BZ}{\cal S}({\rm \textbf{q}},\hbar\omega)d{\rm \textbf{q}}d\hbar\omega}{\int_{\rm BZ}d{\rm \textbf{q}}}=S(S+1).\label{sumrule1}
\end{eqnarray}

We obtain the local dynamic structure factor ${\cal S}(\hbar\omega)$ of DLCB by integrating over momentum transfer along $0.4\leq \rm H\leq0.6$ and $-0.7\leq \rm L\leq-0.3$~[18]. The above equation becomes:
\begin{eqnarray}
\int_{-\infty}^{+\infty}{\cal S}(\hbar\omega)d\hbar\omega=S(S+1).\label{sumrule2}
\end{eqnarray}

Figure 2(d) in the main text shows the background-subtracted energy scan of ${\cal S}(\hbar\omega)$ at \emph{T}=0.25 K and \emph{P}=1.3 GPa in the quantum disordered phase. It includes the correction for the Cu$^{2+}$ magnetic form factor~[19]. In thermal equilibrium, ${\cal S}(\hbar\omega)$ obeys the principle of detailed balance~[20] as:
\begin{eqnarray}
{\cal S}(-\hbar\omega)=e^{-\hbar\omega/k_{\rm B}T}{\cal S}(\hbar\omega).\label{balance}
\end{eqnarray}
At \emph{T}=0.25 K, ${\cal S}(-\hbar\omega)$$\approx0$. Consequently, Eq.~(\ref{sumrule2}) becomes:
\begin{eqnarray}
\int_{0}^{+\infty}{\cal S}(\hbar\omega)d\hbar\omega=S(S+1).\label{sumrule3}
\end{eqnarray}

As there are two well-separated gapped modes including one sharp LM and another broad TM excitation and the contribution from the two-magnon continuum scattering is negligible in DLCB, ${\cal S}(\hbar\omega)$ can be written as ${\cal S}(\hbar\omega)$=${\cal S}^{\rm TM}(\hbar\omega)$+${\cal S}^{\rm LM}(\hbar\omega)$. Because neutron scattering measurements are only sensitive to spin fluctuations perpendicular to the wavevector transfers, the measured intensity can be written as ${\cal I}(\hbar\omega)$=${\cal S}^{\rm TM}(\hbar\omega)(1+\cos^2{\alpha})/2$+${\cal S}^{\rm LM}(\hbar\omega)\sin^2{\alpha}$, where $\alpha$ is the angle between the wavevector transfer and the easy axis. In this particular case $97^\circ$$\leq$$\alpha$$\leq$107$^\circ$ through the \textbf{q}-integrated window, $\cos^2\alpha$$\approx$0 and $\sin^2\alpha$$\approx$1, so ${\cal I}(\hbar\omega)$$\approx$${\cal S}^{\rm TM}(\hbar\omega)/2$+${\cal S}^{\rm LM}(\hbar\omega)$. One then can determine the normalizing constant (${\cal NC}$):
\begin{eqnarray}
{\cal NC}=\frac{{\cal S}^{\rm TM}_{\rm Int}+{\cal S}^{\rm LM}_{\rm Int}}{S(S+1)},\label{NC}
\end{eqnarray}
where ${\cal S}^{\rm TM}_{\rm Int}$ and ${\cal S}^{\rm LM}_{\rm Int}$ are the integrated intensities extracted from Fig.~2(d) in the main text for the TM and LM, respectively.

\section{Quantum Fisher Information}
\begin{figure*}
\includegraphics[width=7cm,bbllx=170,bblly=85,bburx=420,bbury=715,angle=-90,clip=]{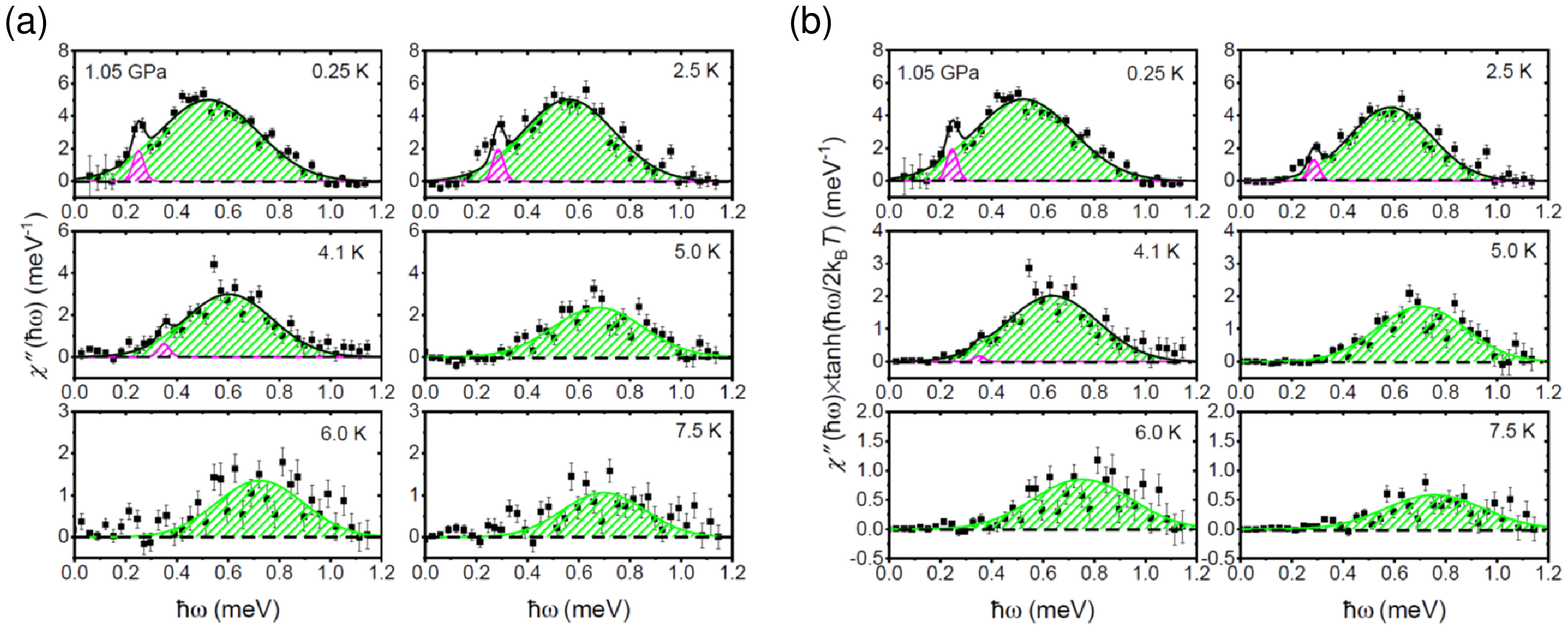}
\caption{Normalized imaginary part of the dynamical spin susceptibility (a) and quantum Fisher information integrand (b) at 1.05 GPa at \emph{T}=0.25 K, 2.5 K, 4.1 K, 5.0 K, 6.0 K, and 7.5 K, respectively. The green and magenta shaded areas represent contributions from the TM and LM, respectively. Note that the LM is traceable up to 4.1 K. The black solid lines at \emph{T}=0.25 K, 2.5 K, and 4.1 K are their sum. The black dashed lines are guides to the eye.}\label{susceptibility}
\end{figure*}
We turn to measure and quantify multipartite entanglement in DLCB, which can provide another insight into emergent states of quantum matter. We apply the quantum Fisher information (QFI)~[21]---a witness for multipartite entanglement through the dynamic susceptibility to magnetic excitation spectra at 1.05 GPa. While this local approach does not directly probe the long-range entanglement, for example, in quantum spin liquids, such measurements can still permit the characterization of exotic quantum phases. The QFI density $f_Q(T)$ can be written as:
\begin{eqnarray}
f_Q(T)=\frac{4}{\pi}\int_{0}^{\infty}d\hbar\omega\tanh(\frac{\hbar\omega}{2k_{\rm B}T})\chi^{''}(\hbar\omega,T),\label{QFI}
\end{eqnarray}
where $\chi^{''}(\hbar\omega)$ is the imaginary part of the dynamic susceptibility. $\chi^{''}(\hbar\omega)$ is related to the local dynamic structure factor ${\cal S}(\hbar\omega)$ via the fluctuation-dissipation theorem~[22],
\begin{eqnarray}
{\cal S}(\hbar\omega)=\frac{1}{\pi}\frac{\chi^{''}(\hbar\omega)}{1-e^{-\hbar\omega/k_{\rm B}T}},\label{fluc}
\end{eqnarray}
${\cal S}(\hbar\omega)$ is the \textbf{q}-integrated dynamic structure factor ${\cal S}(\textbf{q},\hbar\omega)$ as
\begin{eqnarray}
{\cal S}(\hbar\omega)=\int{\cal S}(\textbf{q},\hbar\omega)d\textbf{q},\label{sq}
\end{eqnarray}
which is directly accessible by neutron spectroscopy.

\begin{figure}
\includegraphics[width=8cm,bbllx=25,bblly=10,bburx=600,bbury=480,angle=0,clip=]{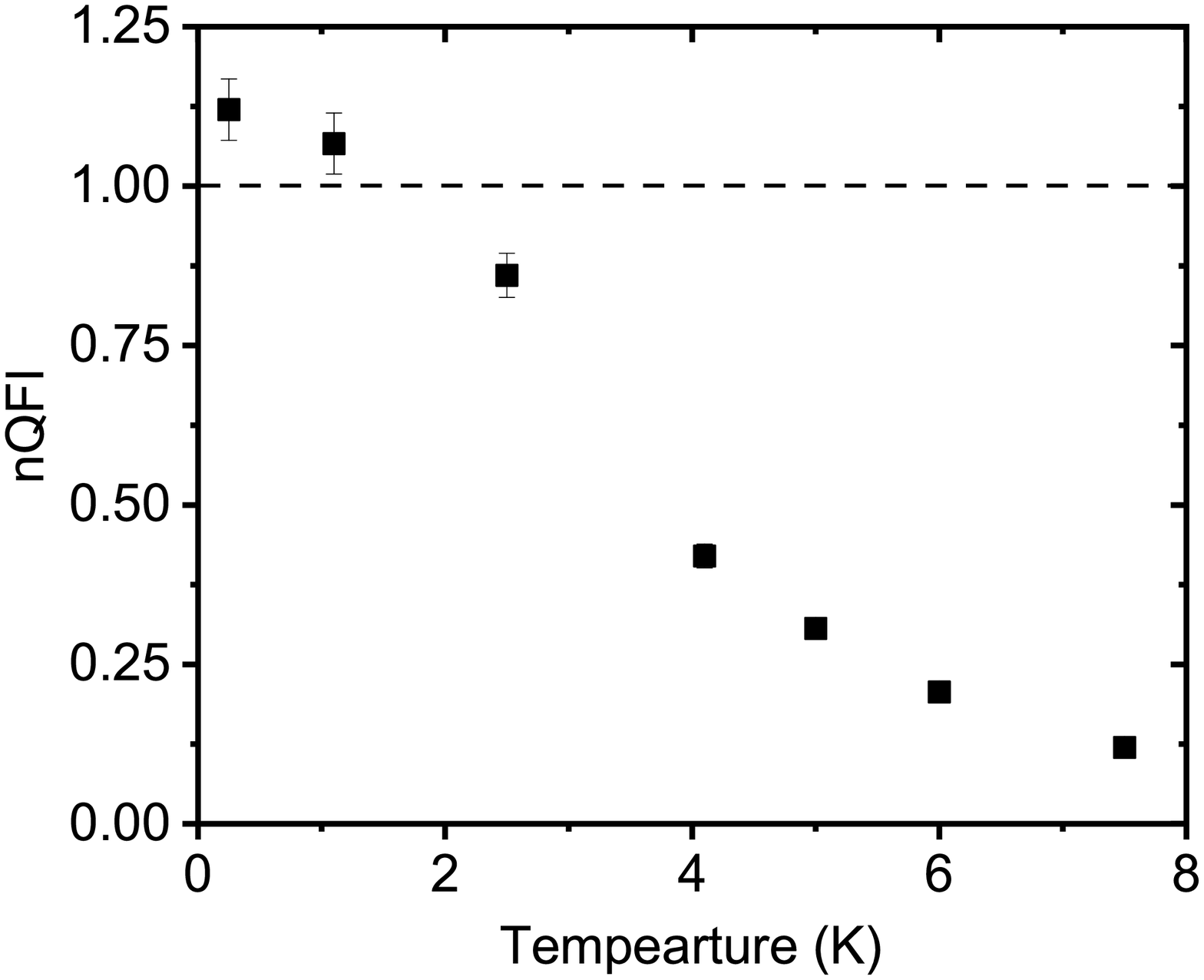}
\caption{Normalized quantum Fisher information as a function of temperature. The system is a bipartite-entangled state entangled in the regime above the dashed line.}\label{fig5}
\end{figure}

Figure~\ref{susceptibility}(a) shows the obtained $\chi^{''}(\hbar\omega)$ using Eq.~(\ref{fluc}) at 1.05 GPa at \emph{T}=0.25 K, 2.5 K, 4.1 K, 5.0 K, 6.0 K, and 7.5 K, respectively. Data were normalized to absolute units by the total moment sum rule. We observe that the energy gap of the TM in the quantum disordered phase slowly grows with temperature, which is driven by a repulsion between thermally excited quasiparticles, and the LM is barely visible above 4 K. Their relevant QFI integrand in Eq.~(\ref{QFI}) is plotted in Fig.~\ref{susceptibility}(b). For neutron scattering studies of spin-\emph{S} systems, the normalized quantum Fisher information nQFI~[23] becomes:
\begin{eqnarray}
{\rm nQFI}(T)=\frac{f_{\rm Q}(T)}{12S^2}>m,\label{nqfi}
\end{eqnarray}
where \emph{S}=1/2 is the spin for DLCB and \emph{m}$>$0 is a divisor of 12$S^2$, then the system is in a state with $\geq$(\emph{m}+1)-partite entangled. The experimentally obtained nQFI using Eq.~(\ref{nqfi}) as a function of temperature at 1.05 GPa is shown in Fig.~\ref{fig5}. The value of nQFI indicates the presence of at least bipartite entanglement up to at least 1.1 K, corresponding to 10$\%$ around of the continuum zone-boundary energy.

\newpage
\section{SUPPLEMENTARY REFERENCES}
[1] F. Awwadi \emph{et al.}, \emph{Strong Rail Spin 1/2 Antiferromagnetic Ladder Systems: (Dimethylammonium)(3,5-Dimethylpyridinium)$\rm CuX_4, X = Cl, Br$}, Inorg.\ Chem. {\bf 47}, 9327 (2008). \newline
[2] G. Ehlers, A. A. Podlesnyak and A. I. Kolesnikov, \emph{The Cold Neutron Chopper Spectrometer at the Spallation Neutron Source—A Review of the First 8 Years of Operation}, Rev. Sci. Instrum. {\bf 87}, 093902 (2016). \newline
[3] B. Winn \emph{et al.}, \emph{Recent Progress on HYSPEC, and Its Polarization Analysis Capabilities}, EPJ Web Conf. {\bf 83}, 093902 (2015).
[4] R. Azuah \emph{et al.,} \emph{DAVE: A comprehensive software suite for the reduction, visualization, and analysis of low energy neutron spectroscopic data.} J. Res. Natl. Inst. Stan. Technol. \textbf{114}, 341 (2009). \newline
[5] D. Schmidiger, P. Bouillot, S. M$\rm\ddot{u}$hlbauer, S. Gvasaliya, C. Kollah, T. Giamarchi, and A. Zheludev, \emph{Spectral and Thermodynamic Properties of a Strong-Leg Quantum Spin Ladder}, Phys. Rev. Lett. {\bf 108}, 167201 (2012). \newline
[6] Y. Ajiro, T. Goto, H. Kikuchi, T. Sakakibara, and T. Inami, \emph{High-Field Magnetization of a Quasi-One-Dimensional S=1 Antiferromagnet Ni(C$_2$H$_8$N$_2$)$_2$NO$_2$(ClO$_4$): Observation of the Haldane Gap}, Phys. Rev. Lett. {\bf 63}, 1424 (1989). \newline
[7] T. Hong, K. P. Schmidt, K. Coester, F. F. Awwadi, M. M. Turnbull, Y. Qiu \emph{et al.}, \emph{Magnetic Ordering Induced by Interladder Coupling in the Spin-1/2 Heisenberg Two-Leg Ladder Antiferromagnet} C$_9$H$_{18}$N$_2$CuBr$_4$, Phys. Rev. B {\bf 89}, 174432 (2014). \newline
[8] T. Hong \emph{et al.}, \emph{Higgs Amplitude Mode in A Two-Dimensional Quantum Antiferromagnet Near the Quantum Critical Point}, Nat. Phys. {\bf 13}, 638 (2017).
[9] S. Toth and B. Lake, \emph{Linear Spin Wave Theory for Single-Q Incommensurate Magnetic Structures}, J. Phys. Condens. Matter {\bf 27}, 166002 (2015). \newline
[10] T. Hong \emph{et al.}, \emph{Evidence for Pressure Induced Unconventional Quantum Criticality in the Coupled Spin Ladder Antiferromagnet} C$_9$H$_{18}$N$_2$CuBr$_4$, Nat. Commun. {\bf 13}, 3073 (2022). \newline
[11] T. Ying, K. P. Schmidt and S. Wessel, \emph{Higgs Mode of Planar Coupled Spin Ladders and Its Observation in} C$_9$H$_{18}$N$_2$CuBr$_4$, Phys. Rev. Lett. {\bf 122}, 127201 (2019). \newline
[12] M. B. Stone \emph{et al.}, \emph{Quasiparticle Breakdown in a Quantum Spin Liquid}, Nature {\bf 440}, 187 (2006). \newline
[13] M. E. Zhitomirsky and A. L. Chernyshev, \emph{Colloquium: Spontaneous Magnon Decays}, Rev. Mod. Phys. {\bf 85}, 219 (2013). \newline
[14] I. Makhfudz, \emph{Fluctuation-Induced First-Order Quantum Phase Transition of The U(1) Spin Liquid in A Pyrochlore Quantum Spin Ice}, Phys. Rev. B {\bf 89}, 024401 (2014).
[15] R. Schaffer, S. Bhattacharjee, and Y.-B. Kim, \emph{Quantum Phase Transition in Heisenberg-Kitaev Model}, Phys. Rev. B {\bf 86}, 224417 (2012). \newline
[16] J.-H. She, J. Zaanen, A. R. Bishop, and A. V. Balatsky, \emph{Stability of Quantum Critical Points in The Presence of Competing Orders}, Phys. Rev. B {\bf 82}, 165128 (2010). \newline
[17] S. Sachdev, \emph{Quantum Phase Transition}, Cambridge Univ. Press, Cambridge, 1999. \newline
[18] The choice of the integration limits is justified by the fact that spectral weights for low-energy magnetic excitations are concentrated in these integral ranges. \newline
[19] S. Shamoto, M. Sato, J. M. Tranquada, B. J. Sternlieb, and G. Shirane, \emph{Neutron-Scattering Study of Antiferromagnetism in $\rm YBa_2Cu_3O_{6.15}$}, Phys. Rev. B {\bf 48}, 13817 (1993). \newline
[20] S. W. Lovesey, \emph{Theory of Neutron Scattering from Condensed Matter}, Clarendon Press, Oxford, (1984). \newline
[21] P. Hauke, M. Heyl, L. Tagliacozzo, and P. Zoller, \emph{Measuring Multipartite Entanglement Through Dynamic Susceptibilities}, Nat. Phys. {\bf 12}, 778 (2016). \newline
[22] W. Marshall and R. D. Lowde, \emph{Magnetic Correlations and Neutron Scattering}, Rep. Prog. Phys. {\bf 31}, 705 (1968). \newline
[23] A. Scheie, P. Laurell, A. M. Samarakoon, B. Lake, S. E. Nagler, G. E. Granroth, S. Okamoto, G. Alvarez, and D. A. Tennant, \emph{Witnessing Entanglement in Quantum Magnets Using Neutron Scattering}, Phys. Rev. B {\bf 103}, 224434 (2021); Phys. Rev. B {\bf 107}, 059902 (2023). \newline

\end{document}